\documentclass[10pt,twocolumn,letterpaper]{article}

\usepackage{cvpr}
\usepackage{times}
\usepackage{epsfig}
\usepackage{graphicx}
\usepackage{amsmath}
\usepackage{amssymb}

\usepackage[breaklinks=true,bookmarks=false]{hyperref}

\cvprfinalcopy 

\def\cvprPaperID{****} 

\begin{document}

\title{OpenRadar:\\
A Toolkit for Prototyping mmWave Radar Applications}


\author{
Arjun Gupta\footnotemark[1]\\
{\tt\small arjung2@illinois.edu}
\and
Dashiell Kosaka\footnotemark[1]\\
{\tt\small dkosaka2@illinois.edu}
\and
Edwin Pan\footnotemark[1]\\
{\tt\small epan2@illinois.edu}
\and
Jingning Tang\footnotemark[1]\\
{\tt\small jtang10@illinois.edu}
\and
Ruihao Yao\footnotemark[1]\\
{\tt\small ruihaoy2@illinois.edu}
\and
Sanjay Patel\\
{\tt\small sjp@illinois.edu}
}
\maketitle
\urlstyle{same}

\footnotetext[1]{Equal contribution. Corresponding email: presenseradar@gmail.com}
\begin{abstract}
Millimeter-Wave (mmWave) radar sensors are gaining popularity for their robust sensing and increasing imaging capabilities. However, current radar signal processing is hardware specific, which makes it impossible to build sensor agnostic solutions.
OpenRadar \cite{openradar} serves as an interface to prototype, research, and benchmark solutions in a modular manner. This enables creating software processing stacks in a way that has not yet been extensively explored. In the wake of increased AI adoption, OpenRadar can accelerate the growth of the combined fields of radar and AI. The OpenRadar API was released on Oct 2, 2019 as an open-source package under the Apache 2.0 license. The code base exists at \url{www.github.com/presenseradar/openradar}.
\end{abstract}

\section{Introduction}
There has been a resurgent interest in radar technology, driven in part by the growth of automotive applications that use radar sensing.  Radar offers significant advantages over other automotive sensing modalities such as LIDAR and standard cameras, which suffer degradation in inclement weather.  Other technological advances are also fueling the adoption of radar, including the use of CMOS technologies, improved on-chip processing power, low cost, and small form factors.  Full solutions are now available that incorporate antennae, radar, and processing on a single device \cite{}.  This growth in capabilities of mmWave radar has opened up a slew of new application domains beyond automotive that benefit from radar sensing, such as people tracking, security, activity monitoring, and robotics.


\renewcommand{\thefootnote}{1} The PreSense\footnote{\url{www.presenseradar.com}} team was formed in the Spring of 2019 as part of the first cohort in the \renewcommand{\thefootnote}{2}Alchemy Foundry\footnote{\url{www.alchemyfoundry.com}} at the University of Illinois at Urbana-Champaign. PreSense focuses on developing new capabilities for emerging multiple-input multiple-output (MIMO) mmWave radar technology, particularly involving advanced signal processing, tracking, machine learning, and sensor fusion.  During the early part of our journey within the Alchemy Foundry, it was evident that radar-agnostic building blocks were missing from this space, and rapid prototyping environments needed to be built from scratch.  We started our journey by building an initial pipeline to produce a 4D point cloud of range, azimuth, elevation, and velocity gathered from a TI radar\cite{}, which we soon thereafter extended to be more customizable, modular, and parameterized to allow us to build trackers, classifiers, and to perform more advanced MIMO techniques to improve resolution.  This was to birth of OpenRadar.



In this paper, we provide an overview of OpenRadar and our guiding philosophies.  Our desire is to create an environment for development of radar solutions that is as valuable as OpenCV is for standard imaging.  In Section 2 of this paper, we provide a brief overview of basic  mmWave radar concepts.  In Section 3, we describe the OpenRadar package, and its structure and implementation. In Section 3 we describe several case studies to illustrate how OpenRadar can facilitate prototyping of radar applications. Section 4 provides some of our thoughts on future roadmap for OpenRadar, and Section 5 provides some similar, parallel works to OpenRadar for reference. Section 6 provides some concluding thoughts.

\section{Background Concepts}

\label{model}
Versatility and robustness is needed in a single API meant to produce results on multiple distinct radar systems. At a low level, there are many forces at work, ranging from the format in which data is captured to the physical design of the radar. For most operations (i.e. algorithms) performed on radar data, these low level specifications can be abstracted away. We have condensed most of these specifications into four different well known elements: frames, (physical/virtual) antennas, chirps, and ADC samples.  

\subsection{Samples}
The lowest level that needs to be dealt with in software is the ADC sample, ADC standing for "analog to digital converter". Every piece of data that a radar receives is initially interpreted as some type of analog signal. When the receiving antennas try to report that piece of data, it needs to be digitized (converted) into some data type a computer can represent. As an example case, 16-bit signed integers are what we used initially that were generated by the TI IWR1642 + DCA1000 radar setup we had.

\subsection{Chirps}
Chirps are one step higher than the ADC sample. At the hardware level, a single chirp is defined by an electromagnetic (EM) wave emitted by a transmit antenna. The back-scatter of this wave is sampled at different times as ADC samples. So, in the context of software, sequential samples taken from these EM waves by receive antenna(s) represent a chirp.

\subsection{Antennas}
There are two types of antennas that can exist in a radar system. Transmit antennas are physical elements in a radar system and are responsible for the transmission of chirps. Physical in this context implies that the antenna has a real size and location. Receive antennas can be seen as physical or virtual. The receive antenna's function is to read in and interpret data so it can then be digitized. Each physical receive antenna, just like their transmitting counterpart, exist as a physical element in a radar system. On the other hand, virtual receivers do not necessarily take up space, yet do have a defined location in the system, and thus are labeled virtual. 



\subsection{Frames}
Analogous to the frames of a camera, radar frames capture the data from a \textit{short slice of time}. The meaning is loose and can encompass any combination of the former elements. Instead of being made up of pixels, the radar frame is made up of varying amount of samples, chirps, and antennas.

\subsection{Range, Doppler, Angle Detections}

\subsection{Multiple-Input Multiple-Output}
The basis of multiple-input multiple-output (MIMO) allows for capturing more data in a efficient manner. In these types of radar systems, we can utilize multiple physical transmit and receive antennas to create a virtual receiver array that captures more rich information about the environment. For the trivial case, assume there is a MIMO array of $M$ physical transmit antennas followed by $N$ physical receive antennas, making a line of antennas. Then the virtual array created would also be a line, but the number of virtual receivers would be $MN$. Each one of these $MN$ virtual receivers captures a time series that inherently differs to some degree from the rest, depending on the spacing of the physical antennas utilized. Combining this data will yield benefits in resolution, noise reduction, and overall usefulness of the data. 

\subsection{FMCW vs. PMCW}
In the previous section about chirps, it was mentioned that a chirp consisted of an emitted EM wave. In the current generations of radar, there are two prominent methods of generating these chirps. Additionally, both these methods emit continuous waves (CW) as opposed to discrete pulses. The end goal of these chirp construction techniques are to help match what is emitted by the transmitters to what is captured by the receivers. These techniques are called frequency modulated continuous wave (FMCW) and phase modulated continuous wave (PMCW), respectively. Just as its name suggests, FMCW sends out a chirp that changes in frequency. This modulation can commonly be seen as a sawtooth waveform in a frequency vs. time graph. PMCW modulates the internal phase encodings of the waves emitted.

\section{OpenRadar Architecture}

While using the package, data is consistently interpreted as Numpy \cite{numpy} arrays for efficient computations including matrix operations. Most of the low level radar concepts are condensed into four basic elements: (ADC) samples, chirps, antennas, frames or scans. Each of these elements are commonly assumed to be organized as their own dimension in an array and operated as such. Brief descriptions of each are as follows.

\begin{itemize}
\item Samples: The lowest level of radar data specifying the value of an analog to digitally converted sample of a EM wave. Each sample provides the magnitude and phase information.
\item Chirps: Emitted continuous EM waves from the radar. Since these signals needs to be discretized, it is seen as groups of samples taken in sequence.
\item Antennas: The physical emitters and receivers used by the radar itself. Combinations of these two can create virtual arrays.
\item Frames: The utilization of any combination and number of the ladder elements. Frames contain snapshots in time analogous to the frames of a video.
\end{itemize}

\subsection{Dataloading}
Radar data is unique in the sense that it doesn't have a uniform structure. Additional logic is required to store and utilize the data. In the OpenRadar package, we created ways to interface with radars and retrieve the data they capture. The raw amount of data transmitted in a scan from a radar can be relatively massive, and these scans that can repeated many times per second. Since the data transmission rates of USB protocols are often not high enough, special interfaces are needed. In some cases, socket connections need to be made for the transmission of user datagram protocol (UDP) packets containing large amounts of data.

\subsection{Range \& Doppler Processing}
Radars were initially designed to take advantage of the oscillating phase of emitted EM waves. Access to this information gives radar superiority in detecting accurate ranges and velocities over many other sensors. The collections of samples and chirps in a frame give range and velocity information, respectively. The rough conversion to the range and velocity domains is done by a performing a fast fourier transform (FFT) \cite{something on FFT?}. However, this operation needs to be done on each group of samples and chirps independently. Moreover, windowing techniques may be desired to eliminate artifacts post-FFT. OpenRadar provides high level functions like \textit{mmwave.dsp.range\_processing()} and \textit{mmwave.dsp.doppler\_processing()} which incorporates these operations into a single function call.

\subsection{Angle of Arrival}
Incorporating MIMO arrays within radars achieves multiple goals. First, the radar captures more data. More importantly, the radar gains the ability to detect the angles of the received signals in the form of yaw and pitch (azimuth and elevataion) relative to the radar. This information is encoded within the slight differences recorded from different physical receivers. By using the assumption of parallel backscatter from an object to a radars receivers, we can see the sampled data is of the form $Ae^{j\phi}[1 e^{jw} e^{j2w} ... e^{jNw}]$ across the antennas. Another FFT can extract the constant phase difference (denoted as $w$) from the signal, which can be seen as an approximate angle of detection. However, this method is only approximate and na\"ive. In contrast, there are many other approaches which are included in OpenRadar that use beamforming or subspace methods to extract more concentrated power spectrums and therefore better angle approximations. Examples include the Capon beamformer, Bartlett beamformer, and MUSIC \cite{some kind of reference on these techniques}.

\subsection{Peak Detection \& Noise Mitigation Tools}
Working with a sensor also means dealing with some degree of noise in the data. In the case of radar data, the signals experience combinations of different types of noise. The OpenRadar package provides various ways of working bypassing this noise to some degree. For certain scenarios, algorithms like constant false alarm rate (CFAR) \cite{something for CFAR} can be used to find outliers (peaks) in the data the have high likelihood of being a real object. In other cases, muffling the noise altogether may be preferred by using filters on the data such as the Log Gabor filter \cite{something for the log gabor} to target specific types of noise present in the data.


\section{Case Studies}

We would like to propose multiple situations in which our package would benefit. As our package evolves, we would want to see the package used in ways that we did not initially think of.

\begin{enumerate}
\item \textbf{Research} has become the primary usage of OpenRadar in the early stages of the package. As we have observed, we are making a bridge from raw radar data to the inputs of AI solutions, most notably deep networks. These research teams take advantage of the high level understanding of the algorithms, allowing the package to execute the low level logic.

\item Since we provide a modular interface, we have high hopes that the package will be used to \textbf{accelerate industrial prototyping}. Similarly to research, this entails that a group may have a specific goal in mind. However, because there are so many design options, the package will allow for quick testing for optimal parameter and algorithm selection. 

\item Although radar is a relatively old technology, the availability of open-source data is lacking. In contrast, LiDAR has been recently widely adopted for the use of autonomous driving, and open-source LiDAR datasets \cite{waymo_open_dataset} are much more abundant as a result. We hope that enabling low-level interfacing with multiple radars will give people the resources to start \textbf{collecting helpful data} which can be turned into public datasets.

\item As the OpenRadar package becomes more mature, we think that it could be a worthwhile \textbf{benchmarking standard}. Analyzing operations, timings, and general performance can become possible through usage.
\end{enumerate}

\section{Roadmap}

Our goal is to be a standard in the radar processing world that follows the architecture of OpenCV. As a result, we would love to improve in the following areas:
\begin{enumerate}
    \item Consistency and reliability of the package and API design.
    \item Bring radar data to machine learning, including basic feature extraction algorithms and our in-house pretrained models.
    \item Have the capability to be run on GPU. 
    \item More algorithms including but not limited to CFAR, AoA and noise removal.
    
\end{enumerate}

\section{Related Work}

We have identified multiple projects following similar structure or topic. We want to briefly introduce these projects and how they display similarities. 

First, regarding the sustainability of OpenRadar as an open-source package. It has been demonstrated with scikit-learn\cite{scikit-learn}, TensorFlow\cite{tensorflow2015-whitepaper}, OpenCV\cite{opencv_library} and PyTorch\cite{paszke2017automatic}.


Lastly, the field of MIMO mmWave FMCW radars are the topic of revolutionary technology. One notable being Google's Soli radar system \cite{wang2016interacting} for recognizing gestures. They successfully designed a radar that easily fit within the chassis of a phone. Their system combined radar DSP of a phone's surroundings to generate input to a convolutional neural network. The result of this is the phone's ability to passively sense user gestures with a non-vision based approach.

\section{Conclusion}

OpenRadar serves as a modular processing library specialized for radars. It has the ability to help solve real world problems efficiently. The generalized structure of the package allows for usage of varying hardware with little to no change. As the technology improves and its research converge, we hope that it will make radar a much more attractive sensing modality. The package has been released as an open-source package and we hope it witnesses wide scale adoption.
\newline

\begin{acknowledgements}
This work is supported by IBM-ILLINOIS Center for Cognitive Computing Systems Research (C3SR) - a research collaboration as part of the IBM AI Horizon Network. The PreSense team wants to thank Prof. Sanjay Patel, without whom this project wouldn't be possible. The team is also grateful for mentorship from Dr. Jinjun Xiong of IBM and Prof. Wen-Mei Hwu. Special thanks to Prof. Haitham Hassanieh, Prof. Minh M. Do and Prof. Erhan Kudeki for the techical advice and Texas Instruments™ for the hardware support.
\end{acknowledgements}

{\small
\bibliographystyle{IEEEtran}
\bibliography{bib_proposal}
}

\end{document}